\documentclass[ptsize 1]{ppmo}    
\usepackage{fancyhdr}
\usepackage{graphicx}                   
\usepackage{natbib}
\bibpunct{(}{)}{;}{a}{}{,}

\def\mbox{\hbox}           

\def\deg{\ifmmode ^\circ                
         \else $^\circ$
         \fi
         \hskip -0.1truecm}
\def\degd#1.#2{                         
               \ifmmode {#1^{\hskip 0.05em\circ}\hskip-0.42em.\hskip0.08em#2}
               \else {#1$^{\hskip 0.05em\circ}\hskip-0.42em.\hskip0.08em$#2}
               \fi
              }
\def\mind#1.#2{                         
               \ifmmode {#1^{\hskip 0.05em\prime}\hskip-0.35em.\hskip0.05em#2}
               \else {#1$^{\hskip 0.05em\prime}\hskip-0.35em.\hskip0.05em$#2}
               \fi
              }
\def\secd#1.#2{                         
               \ifmmode {#1^{\prime\prime}\hskip-0.46em.\hskip0.12em#2}
               \else {#1$^{\prime\prime}\hskip-0.46em.\hskip0.12em$#2}
               \fi
              }
\def\timsecd#1.#2{                      
                  \ifmmode {#1^{\rm s}\hskip-0.39em.\hskip0.08em#2}
                  \else {$#1^{\rm s}\hskip-0.39em.\hskip0.08em#2$}
                  \fi
                 }
\def\hms#1h#2m#3s{                      
                  \relax
                  \ifmmode #1^{\rm h}\,#2^{\rm m}\,#3^{\rm s}
                  \else \hbox{$#1^{\rm h}\,#2^{\rm m}\,#3^{\rm s}$}
                  \fi
                 }
\def\dms#1d#2m#3s{                      
                  \relax
                  \ifmmode #1^\circ\,#2^{\prime}\,#3^{\prime\prime}
                  \else \hbox{$#1^\circ\,#2^{\prime}\,#3^{\prime\prime}$}
                  \fi
                 }
\def\dmsd#1d#2m#3.#4s{                  
                      \relax
                      \ifmmode #1^\circ\,#2^{\prime}\,#3^{\prime\prime}
                               \hskip-0.46em.\hskip0.12em#4
                      \else \hbox{$#1^\circ\,#2^{\prime}\,#3^{\prime\prime}
                            \hskip-0.46em.\hskip0.12em#4$}
                      \fi
                     }
\def\hm#1h#2m{                          
              \relax
              \ifmmode #1^{rm h}\,#2^{\rm m}
              \else \hbox{$#1^{\rm h}\,#2^{\rm m}$}
              \fi
             }
\def\dm#1d#2m{                          
              \relax
              \ifmmode #1^\circ\,#2^{\prime}
              \else \hbox{$#1^\circ\,#2^{\prime}$}
              \fi
             }
\def\hmsd#1h#2m#3.#4s{                  
                      \relax
                      \ifmmode #1^{\rm h}\,#2^{\rm m}\,#3^{\rm s}
                               \hskip-0.39em.\hskip0.08em#4
                      \else \hbox{$#1^{\rm h}\,#2^{\rm m}\,#3^{\rm s}
                            \hskip-0.39em.\hskip0.08em#4$}
                      \fi
                     }
\def\hmd#1h#2.#3m{                  
                  \relax
                  \ifmmode #1^{\rm h}\,#2^{\rm m}
                           \hskip-0.55em.\hskip0.22em#3
                  \else \hbox{$#1^{\rm h}\,#2^{\rm m}
                        \hskip-0.55em.\hskip0.22em#3$}
                  \fi
                 }
\def\mg{\relax                          
        \ifmmode ^{\rm m}
        \else $^{\rm m}$
        \fi
       }
\def\mgd#1.#2{                          
              \relax
              \ifmmode #1^{\rm m}
                       \hskip-0.55em.\hskip0.22em#2
              \else \hbox{#1$^{\rm m}
                    \hskip-0.55em.\hskip0.22em$#2}
              \fi
             }

\def\la{\mathrel{\hbox{\rlap{\hbox{\lower4pt\hbox{$\sim$}}}\hbox{$<$}}}}
\def\ga{\mathrel{\hbox{\rlap{\hbox{\lower4pt\hbox{$\sim$}}}\hbox{$>$}}}}

\def\unitspace{\;}                      

\def\un#1{\ifmmode \unitspace\mbox{\rm #1} 
          \else $\unitspace$#1
          \fi}
\def\pun#1#2{\ifmmode \unitspace\mbox{\rm #1}^{#2} 
             \else $\unitspace$#1$^{#2}$
             \fi}

\def\eV{\un{eV}}                      
\def\Lsun{\ifmmode \un{L}_{\odot}     
          \else $\un{L}_{\odot}$
          \fi}
\def\Msun{\ifmmode \un{M}_{\odot}     
          \else $\un{M}_{\odot}$
          \fi}
\def\mum{\ifmmode \unitspace\mu\mbox{\rm m} 
         \else $\unitspace\mu$m
         \fi}
\def\pc{\un{pc}}                      
\def\sqarcsec{\ifmmode \unitspace\Box''    
              \else $\unitspace\Box''$     
              \fi} 

\def\Bp{\relax                            
        \ifmmode B_{||}                   
        \else $B_{||}$
        \fi}
\def\Bt{\relax                            
        \ifmmode B\!_{\perp}              
        \else $B\!_{\perp}$               
        \fi}
\def\Gcr{\relax                           
         \ifmmode \Gamma\!_{\rm cr}       
         \else $\Gamma\!_{\rm cr}$
         \fi}
\def\ICII{\relax                          
          \ifmmode I_{[\CII]}             
          \else $I_{[\CII]}$
          \fi}
\def\LHtwo{\relax                                 
           \ifmmode L_{\mbox{\rm\scriptsize H}_2} 
           \else $L_{\mbox{\rm\scriptsize H}_2}$  
           \fi}
\def\LIR{\relax                           
         \ifmmode L_{\rm IR}              
         \else $L_{\rm IR}$
         \fi}
\def\LLya{\relax                          
          \ifmmode L_{{\rm Ly}\,\alpha}   
          \else $L_{{\rm Ly}\,\alpha}$
          \fi}
\def\MHtwo{\relax                                 
           \ifmmode M_{\mbox{\rm\scriptsize H}_2} 
           \else $M_{\mbox{\rm\scriptsize H}_2}$  
           \fi}
\def\MHtwodot{\relax                                       
              \ifmmode \dot{M}_{\mbox{\rm\scriptsize H}_2} 
              \else $\dot{M}_{\mbox{\rm\scriptsize H}_2}$  
              \fi}                                         
\def\Mstardot{\relax                      
              \ifmmode \dot{M}_{\ast}     
              \else $\dot{M}_{\ast}$      
              \fi}
\def\nHI{\relax                                      
         \ifmmode n_{\mbox{\scriptsize\rm H\,\sc I}} 
         \else $n_{\mbox{\scriptsize\rm H\,\sc I}}$
         \fi}
\def\nHtwo{\relax                                
           \ifmmode n_{{\mbox{\scriptsize H}}_2} 
           \else $n_{{\mbox{\scriptsize H}}_2}$  
           \fi}
\def\rhostardot{\relax                         
                \ifmmode \dot{\rho}_{\ast}     
                \else $\dot{\rho}_{\ast}$      
                \fi}
\def\rhoZdot{\relax                          
             \ifmmode \dot{\rho}_{\rm Z}     
             \else $\dot{\rho}_{\rm Z}$      
             \fi}

\def\sou#1#2{\relax                       
             \ifmmode {\rm #1}\,{\rm #2}  
             \else #1$\,$#2
             \fi}

\def\NGC#1{\sou{NGC}{#1}}                

\def\Arp#1{\sou{Arp}{#1}}                

\def\qu#1#2{\relax                          
            \ifmmode #1_{\rm #2}            
            \else $#1_{\rm #2}$
            \fi}

\def\CO#1{\ifnum#1=0                    
           \ifmmode \mbox{\rm CO}
           \else {\rm CO}
           \fi
          \else
           \ifnum#1<15
            \ifmmode ^{#1}\mbox{\rm CO}
            \else $^{#1}${\rm CO}
            \fi
           \else
            \ifmmode \mbox{\rm C}^{#1}\mbox{\rm O}
            \else {\rm C}$^{#1}${\rm O}
            \fi
           \fi
          \fi}

\def\COp{\ifmmode \mbox{\rm CO}^+           
         \else {\rm CO}$^+$                 
         \fi}

\def\CS#1{\ifnum#1=0                    
           \ifmmode \mbox{\rm CS}
           \else {\rm CS}
           \fi
          \else
           \ifnum#1<15
            \ifmmode ^{#1}\mbox{\rm CS}
            \else $^{#1}${\rm CS}
            \fi
           \else
            \ifmmode \mbox{\rm C}^{#1}\mbox{\rm S}
            \else {\rm C}$^{#1}${\rm S}
            \fi
           \fi
          \fi}

\def\HCOp{\ifmmode \mbox{\rm HCO}^+          
          \else {\rm HCO}$^+$                
          \fi}
\def\Hthreep{\ifmmode \mbox{\rm H}_3^+         
             \else {\rm H}$_3^+$               
             \fi}

\def\Htwo{\ifmmode \mbox{\rm H}_2              
          \else {\rm H}$_2$                    
          \fi}

\def\HtwoO{\ifmmode \mbox{\rm H}_2\mbox{\rm O} 
           \else {\rm H}$_2${\rm O}            
           \fi}

\def\ion#1#2{\ifmmode \mbox{{\rm #1}}\,\mbox{{\sc #2}} 
        \else {\rm #1}$\,${\sc #2}
        \fi}

\def\rec#1#2{\if#2a                            
              \ifmmode \mbox{{\rm #1}}\alpha   
              \else {\rm #1}$\alpha$
              \fi
             \fi
             \if#2b
              \ifmmode \mbox{{\rm #1}}\beta
              \else {\rm #1}$\beta$
              \fi
             \fi
             \if#2g
              \ifmmode \mbox{{\rm #1}}\gamma
              \else {\rm #1}$\gamma$
              \fi
             \fi}

\newcommand{\tabref}[1]{Table~\protect\ref{#1}}
\newcommand{\figref}[1]{Fig.~\protect\ref{#1}}

\newcommand{\eqref}[1]{Eq.~$\left(\protect\ref{#1}\right)$}

\begin{document}

\pagestyle{fancy}

\title{Extreme Superstarclusters}
   
\volnopage{Vol.0 (2005) No.0, 000--000}  
   
\setcounter{page}{1}      
   
\baselineskip=5mm             

\author{Paul P.~van der Werf
      \mailto{pvdwerf@strw.leidenuniv.nl}
   \and Leonie Snijders}

   \offprints{P.P.~van der Werf}                   

\institute{Leiden Observatory, P.O.~Box~9513,
           NL - 2300~RA~Leiden, The Netherlands\\
             \email{pvdwerf@strw.leidenuniv.nl}
         }


\date{Received~~2006 month day; accepted~~2006~~month day}

\abstract{The presence of superstarclusters is a characteristic feature
  of starburst galaxies. 
  We examine the properties of star forming regions and young star clusters 
  in various environments, ranging from common to extreme. We then
  discuss new high spatial resolution mid-infrared imaging and spectroscopy of
  extreme superstarclusters in the obscured region of the Antennae
  ($\NGC{4038{-}4039}$). We find that the PAH emission in this region is
  not dominated by the superstarclusters but is mostly
  diffuse. Emission line ratios found in our high spatial resolution
  data differ significantly from those in larger apertures, strongly
  affecting the derived results. 
}

\authorrunning{P.P.~van der Werf \& L.~Snijders} 

\titlerunning{Extreme superstarclusters}         

\setlength\baselineskip {5mm}           

\maketitle

%
\section{Scaling starbursts}   
\label{sec.scales}

Starburst galaxies cover an enormous range in luminosity. At the low
luminosity end the small starforming dwarf galaxies such as the Small and Large
Magellanic Clouds have $\LIR=7\cdot10^7\Lsun$ and
$\LIR=7\cdot10^8\Lsun$. More distant infrared-bright 
dwarf galaxies have typically
$\LIR\sim3\cdot10^9\Lsun$. Well-studied nearby starbursts such as 
$\NGC{253}$ and M82 have $\LIR=3\cdot10^{10}\Lsun$ and
$6\cdot10^{10}\Lsun$ \citep[luminosities are from][]{Sandersetal03}. 
At higher luminosities, we have the luminous
infrared galaxies (LIRGs) with $\LIR>10^{11}\Lsun$ 
(e.g., the Antennae, $\NGC{4038{-}4039}$), the ultraluminous
infrared galaxies (ULIRGs) with $\LIR>10^{12}\Lsun$ (e.g., $\Arp{220}$), 
and the hyperluminous
infrared Galaxies (HyLIRGs) with $\LIR>10^{13}\Lsun$. While the luminosity
range spanned is more than five decades, star formation is the
fundamental process in all of these objects. The starbursts that are
most amenable to detailed study are obviously the nearest ones, such as
M82 \citep[for a detailed 
analysis see e.g.,][]{ForsterSchreiberetal01,ForsterSchreiberetal03}. These
are however low or moderate luminosity objects, and it is not trivial to
assess how these objects compare to their higher luminosity, but more
distant relatives. This raises the general question how starbursts of 
different luminosities are related. Two observational facts are directly
relevant to this question:
\begin{enumerate}
\item higher luminosity starburst galaxies have also higher {\it star formation
    efficiencies\/} (SFEs), as measured by the their infrared luminosity per
    unit molecular gas mass: ${\rm SFE}=\LIR/\MHtwo$. So more luminous
    starbursts are also more efficient with their fuel than lower luminosity
    starbursts \citep[][and references therein]{SandersMirabel96}.
\item at the highest luminosities, active galactic nuclei play an
  increasingly important role, both in frequency of occurence and in
  energetic importance \citep[e.g.,][]{Veilleuxetal95,Kimetal98}; 
  this may point to a causal connection between
  extreme starbursts and the formation of supermassive black holes.
\end{enumerate}
In order to put the properties of extreme starbursts into perspective,
it is instructive to compare their SFEs with those of other objects
(see \tabref{tab.SFEs}). It is seen that the SFEs of ULIRGs are
comparable to those of the OMC-1 star forming region. In other words, in ULIRGs
the {\it entire\/} molecular interstellar medium is forming stars at the
same efficiency as the OMC-1 region. If the molecular gas mass is
overestimated by a factor of $~5$ with the standard CO-$\Htwo$
conversion factor, as argued by \citet{Solomonetal97} and
\citet{DownesSolomon98}, the SFEs of ULIRGs even become comparable to
the most active region in Orion: the BN/KL massive star formation core.  

\begin{table}[]
\caption[]{Gas Masses, Infrared Luminosities and SFEs for Various Objects}
\label{tab.SFEs}
\begin{center}
\begin{tabular}{lcccl}
\hline
\noalign{\smallskip}
Object & $\LIR$    & $\qu{M}{gas}$  & SFE                  & References \\
       & [$\Lsun$] & [$\Msun$]      & [$\Lsun \Msun^{-1}$] & \\
\noalign{\smallskip}
\hline
\noalign{\smallskip}
ULIRGs & few $10^{12}$ & few $10^{10}$ & $\sim100$ & 
\citet{Sandersetal91} \\
Milky Way total & $7.4\cdot10^9$ & $4.9\cdot10^9$ & 1.5 & 
\citet{Sodroskietal97} \\
Milky Way molecular clouds & $2.4\cdot10^9$ & $1.3\cdot10^9$ & 1.8 & 
\citet{Sodroskietal97} \\
OMC-1 star forming region & $1.2\cdot10^5$   & $2.2\cdot10^3$ & 54  &
\citet{Ballyetal87}; \\
 & & & & \citet{GenzelStutzki89} \\
Orion BN/KL region & $6\cdot10^4$ & $1.5\cdot10^2$ & 400 & 
\citet{GenzelStutzki89} \\ 
\noalign{\smallskip}
\hline
\end{tabular}
\end{center}
\end{table}

\begin{figure}
\centering
\includegraphics[height=115mm]{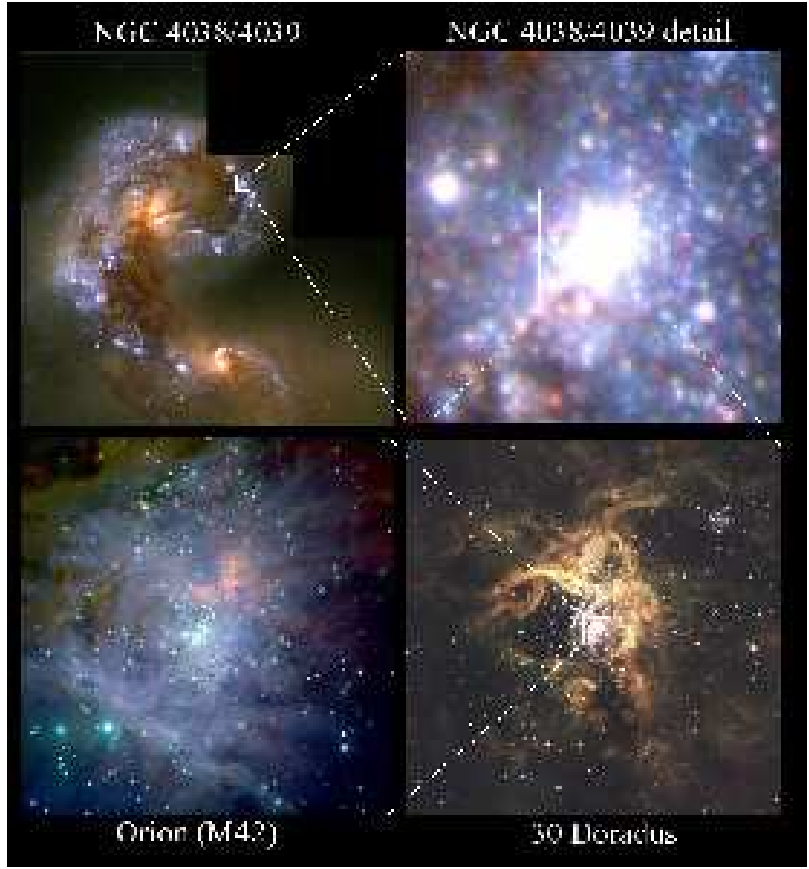}
\caption{Montage illustrating the relative sizes of young starclusters in
  $\NGC{4038{-}4039}$, 30~Doradus and Orion; see \tabref{tab.clusters}
  for an indication of linear sizes.}
\label{fig.clusters}
\end{figure}

\section{Extreme Superstarclusters}
\label{sec.superstarclusters}

The question now arises whether young star clusters in starburst
galaxies of various luminosities also differ in properties. A
characteristic feature of starburst galaxies is the presence of 
{\it superstarclusters\/}, luminous and compact clusters of young stars.
It is still a matter of debate whether all stars in starbursts are born
in superstarclusters. For instance, \citet{Meureretal95} estimated that
only 20\% of the ultraviolet (UV) emission from starbursts comes from young
compact clusters. This result will certainly be incorrect if
young superstarclusters are more obscured than the field population,
which is likely the case. A more likely scenario is then that star
formation in superstarclusters is the dominant mode in starburst
galaxies, and the dissolving clusters give rise to a more diffuse 
population of young stars \citep[e.g.,][]{Mengeletal05}.\\

A compilation of typical properties of star clusters ranging from the
extreme to the common is given in \figref{fig.clusters} and
\tabref{tab.clusters}.
In this table, $d$ is the total linear size of the cluster as seen
in \figref{fig.clusters}. $M_*$ is the stellar mass 
within a fixed radius of $4.5\pc$ (in the case of [WS99]2, this is a
dynamical mass; the other masses are derived from star counts). $\rho_*$
is the corresponding average stellar mass density within this region,
and $\qu{n}{equiv}$ is the equivalent number density of hydrogen atoms.
It is seen that the most extreme superstarclusters are larger in linear
dimension, more massive, and have densities similar to those of the
cores of star forming molecular clouds, indicating a highly 
efficient conversion of mass into stars in the most extreme objects.

\begin{table}[]
\caption[]{Properties of Young Star Clusters in $\NGC{4038{-}4039}$ ([W99]2),
  30~Doradus (R136) and Orion (M42)} 
\label{tab.clusters}
\begin{center}
\begin{tabular}{lccccl}
\hline
\noalign{\smallskip}
Object & $d$ & $M_*$     & $\rho_*$  & $\qu{n}{equiv}$  & References \\
       & [pc] & [$\Msun$] & [$\Msun\pun{pc}{3}$] & [${\rm cm}^{-3}$] & \\
\noalign{\smallskip}
\hline
\noalign{\smallskip}
[W99]2 & 100 & $2\cdot10^6$ & 5200 & $1.6\cdot10^5$ & \citet{Mengeletal02} \\
R136   &  10 & $3\cdot10^4$ & 80 & 2500 & \citet{Brandletal96} \\
M42    &   2 & 1800 & 5 & 200 & \citet{HillenbrandHartmann98} \\ 
\noalign{\smallskip}
\hline
\end{tabular}
\end{center}
\end{table}

\section{Mid-infrared Observations of the Antennae Overlap Region}
\label{sec.Antennae}

Since stars form in dusty molecular clouds, we may expect young
superstarclusters to be optically obscured. This is illustrated
dramatically by the Antennae ($\NGC{4038{-}4039}$), where the bolometric
luminosity is not dominated by the optically visible star
clusters but by a visually obscured, dusty region where the two disks
overlap (a region first highlighted as an active star formation site
with VLA observations by \citet{HummelVanDerHulst86}).  The most
luminous cluster in this region produces 15\% of the total $15\mum$
luminosity of the entire Antennae system \citep{Mirabeletal98}.

Such superstarclusters are of interest as potentially the youngest
simple coeval stellar populations in starbursts and thus furnish
excellent tests for the properties of the most massive stars formed in
these systems. For sufficiently massive and young superstarclusters,
they may offer the opportunity of directly measuring a possible upper
mass cutoff of the stellar Initial Mass Function (IMF). Mid-infrared
nebular fine-structure lines are excellent probes of such systems, since
they are relatively unaffected by dust and can be used to measure the
temperature of the ionizing radiation field, and hence the mass of the
most massive stars present.

\begin{figure}
\centering
\includegraphics[height=110mm]{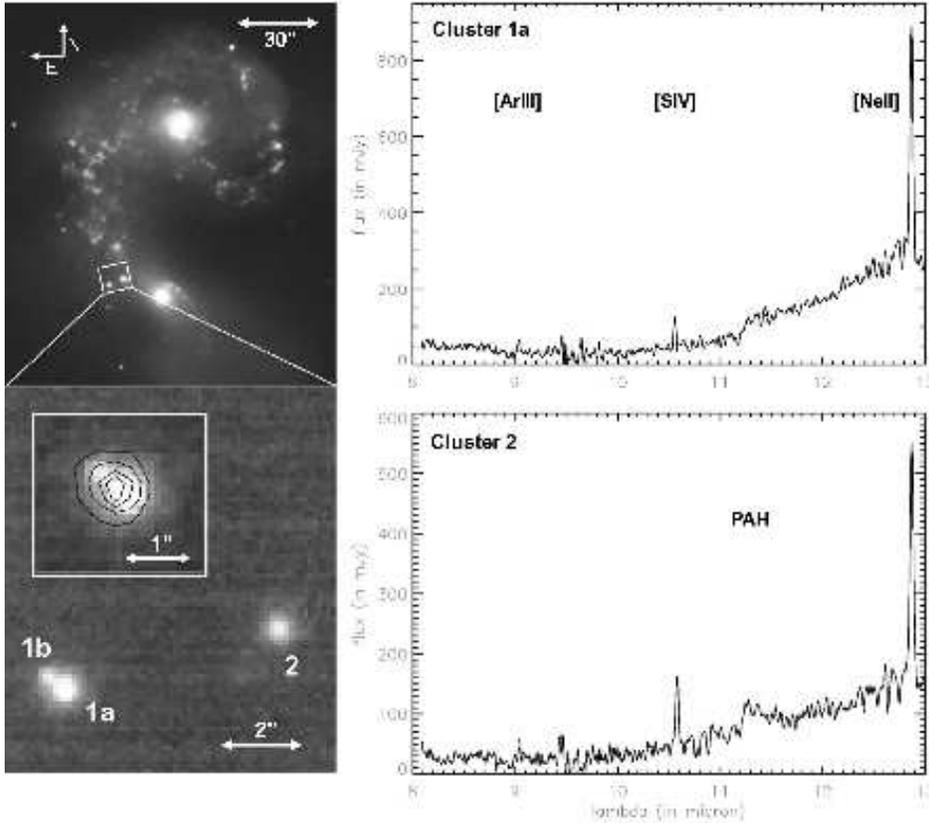}
\caption{The VISIR image of the Antennae overlap region in the
  $12.8\mum$ [$\ion{Ne}{ii}$] filter is shown in the lower left. The
  inset shows a blow-up of the clusters 1a and 1b, with contours of the
  emission in the $11.3\mum$ PAH filter overlaid. 
  The field observed is indicated in the
  $\qu{K}{s}$ image on the upper left. Integrated VISIR spectra of two of the
  clusters are shown on the right-hand side.}
\label{fig.Antennae}
\end{figure}

We have recently used VISIR on the Very Large Telescope of the European
Southern Observatory at Paranal (Chile) to obtained $N$-band ($8-13\mum$)
data of the two most luminous superstarclusters in the overlap region in
the Antennae. Our dataset consists of $\secd 0.3$ resolution
imaging in a number of narrow-band
filters in the $N$-band, and long-slit spectroscopy with a $\secd 0.75$ slit.
Some key results are shown in \figref{fig.Antennae}. 

Inspection of \figref{fig.Antennae} reveals a few surprising results:
\begin{enumerate}
\item the Eastern cluster is separated into two components, separated by
  approximately $\secd 0.5$ ($50\pc$); the brightest of these two
  (cluster~1a) is slightly resolved; this is cluster [WS95]80 of
  \citet{WhitmoreSchweizer95};
\item cluster~1b has no counterpart in any other available dataset; we
  derive a visual extinction $A_V>65\mg$ towards this cluster;
\item remarkably, the $11.3\mum$ emission shows a different morphology,
  suggesting a common envelope of emission from hot dust and polycyclic
  aromatic hydrocarbons (PAHs);
\item comparison with Spitzer-IRS spectra (Brandl, priv.~comm.) 
  with a $5''$ slit reveals
  that $\ga50\%$ of the $12\mum$ continuum is detected in the
  $\secd 0.75$ VISIR slit; however, the equivalent width of the $11.3\mum$
  PAH feature in the VISIR data is much smaller than in the larger
  aperture Spitzer spectra; 
\item both clusters exhibit emission in the $10.5\mum$ [$\ion{S}{iv}$]
  line, an ionization stage requiring $34.8\eV$ (while the $12.8\mum$
  [$\ion{Ne}{ii}$] lines requires only $21.6\eV$); in particular in
  cluster~2 the [$\ion{S}{iv}$]/[$\ion{Ne}{ii}$] ratio in our data is
  higher than in larger aperture Spitzer data, significantly affecting
  the interpretation of the results.
\end{enumerate}

The low equivalent width of the PAH emission indicates that the PAH
emission is not preferentially excited by the superstarclusters, but is
dominated by more diffuse emission, excited by the softer UV radiation
from more widespread young stars of slightly later type. The PAH
emission is therefore not a good tracer of the most recent star
formation, and may provide a better measure of the star formation integrated
over a somewhat longer timescale. 

A simple analysis of the [$\ion{S}{iv}$]/[$\ion{Ne}{ii}$] ratio in cluster~1a
indicates a radiation field corresponding to an O3 star; comparison with
the total luminosity of this cluster then requires approximately 1000 of
such stars to be present. This result may reveal a real absence of 
stars more massive than O3; alternatively, earlier spectral
types could still be present if they are formed in ultracompact
$\ion{H}{ii}$ regions, of which the emission lines are strongly quenched.
A more detailed analysis of these results is in preparation (Snijders et al.).

\begin{acknowledgements}
It is a pleasure to thank the organisers for this thoroughly enjoyable meeting.
\end{acknowledgements}

\bibliographystyle{apj}
\bibliography{%
strings,%
LMCSMC,%
M82,%
MilkyWay,%
NGC4038-4039,%
OrionA,%
starbursts,%
ULIRGs%
}

\begin{thebibliography}{20}
\expandafter\ifx\csname natexlab\endcsname\relax\def\natexlab#1{#1}\fi

\bibitem[{{Bally} {et~al.}(1987){Bally}, {Stark}, {Wilson}, \&
  {Langer}}]{Ballyetal87}
{Bally}, J., {Stark}, A.~A., {Wilson}, R.~W., \& {Langer}, W.~D. 1987, \apjl,
  312, L45

\bibitem[{{Brandl} {et~al.}(1996){Brandl}, {Sams}, {Bertoldi}, {Eckart},
  {Genzel}, {Drapatz}, {Hofmann}, {Loewe}, \& {Quirrenbach}}]{Brandletal96}
{Brandl}, B., {Sams}, B.~J., {Bertoldi}, F., {Eckart}, A., {Genzel}, R.,
  {Drapatz}, S., {Hofmann}, R., {Loewe}, M., \& {Quirrenbach}, A. 1996, \apj,
  466, 254

\bibitem[{Downes \& Solomon(1998)}]{DownesSolomon98}
Downes, D. \& Solomon, P.~M. 1998, ApJ, 507, 615

\bibitem[{{F{\" o}rster Schreiber} {et~al.}(2001){F{\" o}rster Schreiber},
  {Genzel}, {Lutz}, {Kunze}, \& {Sternberg}}]{ForsterSchreiberetal01}
{F{\" o}rster Schreiber}, N.~M., {Genzel}, R., {Lutz}, D., {Kunze}, D., \&
  {Sternberg}, A. 2001, ApJ, 552, 544

\bibitem[{{F{\" o}rster Schreiber} {et~al.}(2003){F{\" o}rster Schreiber},
  {Genzel}, {Lutz}, \& {Sternberg}}]{ForsterSchreiberetal03}
{F{\" o}rster Schreiber}, N.~M., {Genzel}, R., {Lutz}, D., \& {Sternberg}, A.
  2003, ApJ, 599, 193

\bibitem[{Genzel \& Stutzki(1989)}]{GenzelStutzki89}
Genzel, R. \& Stutzki, J. 1989, ARA\&A, 27, 41

\bibitem[{{Hillenbrand} \& {Hartmann}(1998)}]{HillenbrandHartmann98}
{Hillenbrand}, L.~A. \& {Hartmann}, L.~W. 1998, \apj, 492, 540

\bibitem[{{Hummel} \& {van der Hulst}(1986)}]{HummelVanDerHulst86}
{Hummel}, E. \& {van der Hulst}, J.~M. 1986, \aap, 155, 151

\bibitem[{{Kim} {et~al.}(1998){Kim}, {Veilleux}, \& {Sanders}}]{Kimetal98}
{Kim}, D.-C., {Veilleux}, S., \& {Sanders}, D.~B. 1998, \apj, 508, 627

\bibitem[{{Mengel} {et~al.}(2002){Mengel}, {Lehnert}, {Thatte}, \&
  {Genzel}}]{Mengeletal02}
{Mengel}, S., {Lehnert}, M.~D., {Thatte}, N., \& {Genzel}, R. 2002, \aap, 383,
  137

\bibitem[{{Mengel} {et~al.}(2005){Mengel}, {Lehnert}, {Thatte}, \&
  {Genzel}}]{Mengeletal05}
---. 2005, A\&A, 443, 41

\bibitem[{{Meurer} {et~al.}(1995){Meurer}, {Heckman}, {Leitherer}, {Kinney},
  {Robert}, \& {Garnett}}]{Meureretal95}
{Meurer}, G.~R., {Heckman}, T.~M., {Leitherer}, C., {Kinney}, A., {Robert}, C.,
  \& {Garnett}, D.~R. 1995, \aj, 110, 2665

\bibitem[{Mirabel {et~al.}(1998)Mirabel, Vigroux, Charmandaris, Sauvage,
  Gallais, Tran, Cesarsky, Madden, \& Duc}]{Mirabeletal98}
Mirabel, I.~F., Vigroux, L., Charmandaris, V., Sauvage, M., Gallais, P., Tran,
  D., Cesarsky, C., Madden, S.~C., \& Duc, P.-A. 1998, A\&A, 333, L1

\bibitem[{Sanders {et~al.}(2003)Sanders, Mazzarella, Kim, Surace, \&
  Soifer}]{Sandersetal03}
Sanders, D.~B., Mazzarella, J.~M., Kim, D.-C., Surace, J., \& Soifer, B.~T.
  2003, ApJS, 126, 1607

\bibitem[{Sanders \& Mirabel(1996)}]{SandersMirabel96}
Sanders, D.~B. \& Mirabel, I.~F. 1996, ARA\&A, 34, 749

\bibitem[{Sanders {et~al.}(1991)Sanders, Scoville, \& Soifer}]{Sandersetal91}
Sanders, D.~B., Scoville, N.~Z., \& Soifer, B.~T. 1991, ApJ, 370, 158

\bibitem[{{Sodroski} {et~al.}(1997){Sodroski}, {Odegard}, {Arendt}, {Dwek},
  {Weiland}, {Hauser}, \& {Kelsall}}]{Sodroskietal97}
{Sodroski}, T.~J., {Odegard}, N., {Arendt}, R.~G., {Dwek}, E., {Weiland},
  J.~L., {Hauser}, M.~G., \& {Kelsall}, T. 1997, \apj, 480, 173

\bibitem[{Solomon {et~al.}(1997)Solomon, Downes, Radford, \&
  Barrett}]{Solomonetal97}
Solomon, P.~M., Downes, D., Radford, S. J.~E., \& Barrett, J.~W. 1997, ApJ,
  478, 144

\bibitem[{Veilleux {et~al.}(1995)Veilleux, Kim, Sanders, Mazzarella, \&
  Soifer}]{Veilleuxetal95}
Veilleux, S., Kim, D.-C., Sanders, D.~B., Mazzarella, J.~M., \& Soifer, B.~T.
  1995, ApJS, 98, 171

\bibitem[{{Whitmore} \& {Schweizer}(1995)}]{WhitmoreSchweizer95}
{Whitmore}, B.~C. \& {Schweizer}, F. 1995, AJ, 109, 960

\end{thebibliography}



\label{lastpage}

\end{document}